\documentclass[12pt,journal]{IEEEtran}
\usepackage{amssymb}
\usepackage{amsmath}
\usepackage[english]{babel}
\usepackage{url}
\usepackage{amsfonts}
\usepackage{graphicx}
\usepackage{epstopdf}
\usepackage{rotating}

\makeatletter
\def\markboth#1#2{\def\leftmark{\@IEEEcompsoconly{\sffamily}\MakeUppercase{\protect#1}}%
\def\rightmark{\@IEEEcompsoconly{\sffamily}\MakeUppercase{\protect#2}}}
\makeatother

\providecommand{\tabularnewline}{\\}

\begin{document}
\title{Statistical Characterization and Mitigation of NLOS Bias in UWB Localization Systems}
\author{Francesco~Montorsi, Fabrizio Pancaldi, and~Giorgio~Matteo~Vitetta
\thanks{F.~Montorsi and G.~M.~Vitetta are with Department of Information Engineering, University of Modena e Reggio Emilia (e-mail: francesco.montorsi@unimore.it and giorgio.vitetta@unimore.it).
            F. Pancaldi is with Department of Science and Methods for Engineering, University of Modena and Reggio Emilia (email: fabrizio.pancaldi@unimore.it).}}
\maketitle

\begin{abstract}
In this paper the problem of the joint statistical characterization
of the NLOS bias and of the most representative features of LOS/NLOS
UWB waveforms is investigated. In addition, the performance of
various \emph{maximum-likelihood} (ML) estimators for joint
localization and NLOS bias mitigation is assessed. Our numerical
results evidence that: a) the accuracy of all the considered
estimators is appreciably affected by the LOS/NLOS conditions of the
propagation environment; b) a statistical knowledge of multiple
signal features can be exploited to mitigate the NLOS bias, so
reducing the overall localization error.
\end{abstract}


\IEEEpeerreviewmaketitle

\markboth{Advances in Electronics and Telecommunications, accepted for publication}{F. Montorsi, F. Pancaldi and G.M. Vitetta: Statistical Characterization and Mitigation of NLOS Bias in UWB Localization Systems}

\section{Introduction\label{sec:Introduction}}

Wireless localization in harsh communication environments (e.g., in
a building where wireless nodes are separated by concrete walls and
other obstacles) can be appreciably affected by direct path attenuation
and NLOS conditions. In principle, the effects of NLOS errors can
be mitigated adopting techniques for detecting LOS/NLOS conditions
or algorithms for estimating the errors themselves (i.e., the \emph{bias}
due to obstacles). These approaches can potentially improve the overall
accuracy and are expected to satisfy certain requirements; in particular,
they should be \emph{robust} against changes in the propagation environment,
\emph{stable} (i.e., they should never amplify NLOS errors) and should
be able to exploit all the available data (e.g., received waveforms
and \emph{a priori} information).

Various solutions for NLOS error mitigation in UWB environments are
available in the technical literature \cite{wed_model},
\cite{nlos_marano}, \cite{decarli}, \cite{venkatesh_buehrer}. In
particular, a simple deterministic model, dubbed \emph{wall extra
delay}, is proposed in \cite{wed_model} to estimate the bias
introduced by walls. A non-parametric support vector machine is
employed in \cite{nlos_marano} for \emph{joint} bias mitigation and
channel status detection; this approach exploits multiple features
extracted from received signals in a non-statistical fashion. A few
classification algorithms for LOS/NLOS detection are compared in
\cite{decarli}, where it is shown that the best solution is offered
by a statistical strategy based on the joint \emph{probability
density function} (pdf) of the delay spread and the kurtosis
extracted from the received signals. Finally, in
\cite{venkatesh_buehrer} statistical models for the \emph{time of
arrival} (TOA), the \emph{received signal strength} (RSS) and the
\emph{root mean square delay spread} (RDS) are developed and an
iterative estimator for bias mitigation is devised.

The contribution of this paper is twofold. In fact, first of all,
the problem of joint statistical modeling of multiple features
extracted from a database of waveforms acquired in a TOA-based
localization system is investigated. Note that, as far as we know,
in the technical literature only univariate models for bias
mitigation have been proposed until now (e.g., see \cite{wed_model},
\cite{venkatesh_buehrer} and \cite{zhang_tsuboi}). The use of
multiple signal features in UWB localization systems has been
investigated in \cite{decarli} for channel state detection only and
in \cite{guvenc}, where, however, the considered features (namely,
the kurtosis, the mean excess delay and the delay spread) have been
modelled as independent random variables. The second contribution
offered by this manuscript is represented by a performance
comparison of various \emph{maximum-likelihood} (ML) estimators for
TOA-based localization. In particular, unlike other papers (e.g.,
see \cite{nlos_marano}, \cite{decarli}) we illustrate some numerical
results referring to the accuracy of different localization
strategies, rather than to capability of the bias removal on a
single radio link.

The remaining part of this paper is organized as follows. In Section
\ref{sec:Experimental-activities} some information about our UWB
experimental campaign and about the features extracted from the acquired
data are provided. In Section \ref{sec:Localization-algorithms} some
estimation algorithms for UWB radiolocalization are described, whereas
their performance is compared in Section \ref{sec:Numerical-results}.
Finally, some conclusions are given in Section \ref{sec:Conclusions}.

\section{Experimental setup\label{sec:Experimental-activities}}

\subsection{Measurement arrangement}

A measurement campaign was conducted by our research group in the
second half of 2010 in order to assess the impact of NLOS bias errors
on a real world UWB system for localization; all the measured data
were acquired by means of two FCC-compliant PulsON220 radios commercialised
by TimeDomain and were collected in a database. Such devices are equipped
with omnidirectional antennas, are characterized by a -10 dB bandwidth
and a central frequency equal to 3.2 GHz and 4.7 GHz, respectively,
and perform two-way TOA ranging; they also allow to store digitised
received waveforms (a sampling frequency of 24.2 GHz and 14 bits per
sample are used).

\begin{figure}
\begin{centering}
\includegraphics[width=8cm]{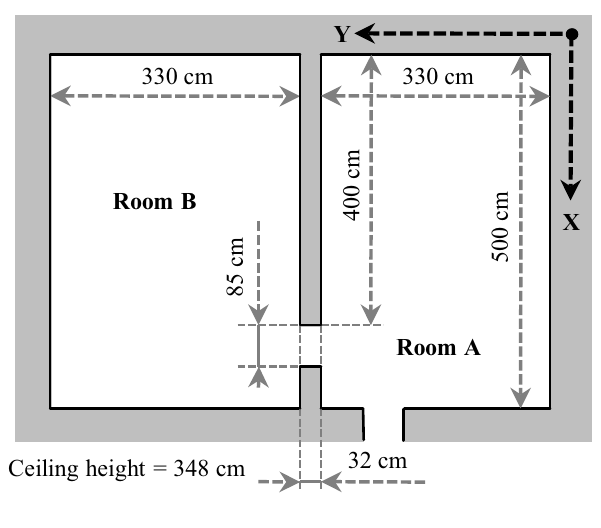}
\par\end{centering}

\caption{Map of the enviroment of our experimental campaign.\label{fig:environment}}
\end{figure}

It is worth mentioning that various databases providing a collection
of sampled UWB waveforms acquired in experimental campaigns and
useful for assessing the performance of localization algorithms are
already available (e.g., see \cite{wprb_database},
\cite{usc_database}). However, our database has been specifically
generated to assess the correlation between the NLOS bias error and
various features extracted from the received signals, as it will
become clearer in the next Paragraph. Our measurement campaign
consisted of two phases. First, the transmitter was placed in a
given room (room A in Fig. \ref{fig:environment}) and the receiver
in an adjacent room (room B in Fig. \ref{fig:environment}) separated
from room A by a wall having thickness $t_{wall}=32$ cm (NLOS
condition); in addition, the transmit antenna was kept fixed,
whereas the receive antenna was placed in
$N_{acq}^{\text{\tiny{NLOS}}}=174$ distinct vertices of a dense
square grid (the distance between a couple of nearest vertices was
equal to 21 cm). In the second phase of our measurement campaign
both the transmitter and the receiver were placed in room B; then,
the transmit antenna was kept fixed, whereas the receive antenna was
moved on the same dense grid as in the first phase to acquire the
UWB waveform in $N_{acq}^{\text{\tiny{LOS}}}=105$ distinct vertices.
The positions of the transmit antenna in the two phases were
$(0.26,2.97)$ m and $(0.26,3.97)$ m, respectively, if the coordinate
system shown in Fig. \ref{fig:environment} is employed. The distance
between the two antennas varied between 1 m and about 5 m; larger
distances were not taken into consideration since we were interested
in indoor ranging only. It is important that to note that the choice
of the measurement scenarios described above is motivated by the
fact that UWB signals experienced similar propagation in both
phases.

For each position of the receive antenna up to 25 realizations of
the received UWB signal were acquired (each one lasting $T_{acq}=110$
ns), in order to filter out the effects of the measurement noise and
the TOA estimate outliers in the postprocessing phase. The analysis
of the multiple waveforms referring to each link evidenced the presence
of few time-variant artifacts (mostly due to the movements of the
people behind the radios); as a matter of fact, the channel in our
measurement campaign can be deemed quasi-static and is characterized
by reflections mainly due to the walls, the ceiling and the floor
of the room (or rooms) hosting the radio devices.

Finally, for each acquired waveform, a TOA estimate (evaluated by
the PulsON220 devices) and the actual transmitter-receiver distance
(evaluated by means of a metric tape with an accuracy better than
1 cm) were also stored in the database. The TOA estimates have been
used to provide a common time frame to all the acquired waveforms;
this has made possible the estimation of the mean excess delay and
of other signal statistics in the signal processing phase.

\subsection{Statistical modeling of signal features\label{sub:Statistical-results}}

In each of the two measurement phases described in the previous
Paragraph, the procedure for acquiring UWB waveforms worked as
follows. First, the UWB radios carried out their handshaking
procedure for achieving mutual time synchronization; then, the
receiver sampled the incoming waveforms and generated a TOA estimate
for each of them; let $r_{i}^{\text{\tiny{LOS}}}(t)$
($r_{i}^{\text{\tiny{NLOS}}}(t)$) and
$\tau_{i}^{\text{\tiny{LOS}}}\triangleq\mathrm{TOA}(r_{i}^{\text{\tiny{LOS}}}(t))$
($\tau_{i}^{\text{\tiny{NLOS}}}\triangleq\mathrm{TOA}(r_{i}^{\text{\tiny{NLOS}}}(t))$)
denote the received waveform and the corresponding TOA\ estimate
(here the function $\mathrm{TOA}(\cdot)$ represents the internal
procedure adopted by the PulsON220 devices, based on energy
thresholds and a go-back technique, to estimate the TOA; see
\cite{toa_estimators} for further details) acquired over the $i$-th
link in LOS (NLOS)
conditions. The model%
\footnote{The superscripts LOS\ and NLOS are not explicitly indicated in the
following expressions, when not strictly needed, to ease the reading.%
}
\begin{equation}
\tau_{i}=\frac{d_{i}}{c_{0}}+b_{i}+w_{i}\label{eq:experimental_sigmodel}
\end{equation}
was adopted for the measured TOA (for both LOS and NLOS conditions).
Here, $c_{0}$ is the speed of light, $d_{i}$ denotes the \emph{distance}
between the transmitter and the receiver, $b_{i}$ is the NLOS \emph{bias}
(in seconds) affecting the TOA measurement (the values taken on by
this random parameter are always positive for NLOS links and null
for LOS links%
\footnote{In this case the \emph{probability density function} (pdf) of $b_{i}$
is $f_{b}(b)=\delta(b)$.%
}) and $w_{i}\sim\mathcal{N}(0,\sigma_{w,i}^{2})$ is the \emph{measurement
noise}; in addition, the expression $\sigma_{w,i}^{2}=\gamma\sigma_{n}^{2}d_{i}^{\beta}$
is adopted for the variance of the measurement noise, where $\gamma$
is a parameter depending on both the specific TOA estimator employed
in the ranging measurements and on various parameters of the physical
layer, and $\beta$ is the \emph{path-loss exponent} (a known and
fixed value is assumed for this parameter in both LOS and NLOS conditions
\cite{venkatesh_buehrer}).
\begin{table*}
\begin{centering}
\begin{tabular}{|c|c|c|c|c|c|c|c|}
\hline
\begin{tabular}{c}
Correlation\tabularnewline
with $b_{i}$ \tabularnewline
\end{tabular}  & $x_{i,0}$ & $x_{i,1}$ & $x_{i,2}$ & $x_{i,3}$ & $x_{i,4}$ & $x_{i,5}$ & $d_{i}$\tabularnewline
\hline
NLOS case & 0.795 & 0.852 & 0.894 & 0.641 & 0.454 & 0.644 & 0.609\tabularnewline
\hline
LOS case & 0.602 & 0.586 & 0.129 & 0.666 & 0.586 & 0.119 & 0.629\tabularnewline
\hline
\end{tabular}
\par\end{centering}

\caption{Absolute values of correlation coefficients between $b_{i}$ and $x_{i,j}$,
with $j=0,....,N_{f}-1$.\label{tab:corr_values_b_xj}}
\end{table*}

In the following we focus on the problem of estimating the bias $b_{i}$
(affecting the TOA estimate $\tau_{i}$ (\ref{eq:experimental_sigmodel}))
from a set of $N_{f}=6$ different {}``features\textquotedblright{}
$\left\{ x_{i,j}\text{, }j=0\text{, }1\text{, }...\text{, }5\right\} $
extracted from the received waveform $r_{i}(t)$. In particular, like
in \cite{nlos_marano}, the following features were evaluated for
the set of the received waveforms:
\begin{enumerate}
\item the \emph{maximum signal amplitude} $x_{i,0}=r_{max,i}\triangleq\max_{t}|r_{i}(t)|$;
\item the \emph{mean excess delay} $x_{i,1}=\tau_{m,i}\triangleq\int_{0}^{\infty}t\frac{|r_{i}(t)|^{2}}{\varepsilon_{i}}dt$
(the parameter $\varepsilon_{i}$ is defined below);
\item the \emph{delay spread }$x_{i,2}=\tau_{ds,i}\triangleq\int_{0}^{\infty}(t-\tau_{m,i})^{2}\frac{|r_{i}(t)|^{2}}{\varepsilon_{i}}dt$;
\item the \emph{energy} $x_{i,3}=\varepsilon_{i}\triangleq\int_{0}^{\infty}|r_{i}(t)|^{2}dt$;
\item the \emph{rise time $x_{i,4}=t_{rise,i}\triangleq\min\left\{ t:|r_{i}(t)|/\max_{t}|r_{i}(t)|>0.9\right\} -\min$
}$\left\{ t:|r_{i}(t)|/\max_{t}|r_{i}(t)|>0.1\right\} $;
\item the \emph{kurtosis} $x_{i,5}=\kappa_{i}\triangleq\frac{1}{\sigma_{|r|}^{4}T}\int_{T}\left(|r_{i}(t)|-\mu_{|r|}\right)^{4}dt$,
where $\mu_{|r|}\triangleq\frac{1}{T}$ $\int_{T}|r_{i}(t)|dt$, $\sigma_{|r|}^{2}\triangleq\frac{1}{T}\int_{T}(|r_{i}(t)|-\mu_{|r|})^{2}dt$
and $T$ denotes the \emph{observation time}.
\end{enumerate}
In the following the set $\left\{ r_{i}(t)\right\} $ of received
waveforms is modelled as a random process, so that the above mentioned
features form a set of correlated random variables; in addition, all
of them are statistically correlated with the TOA bias. The last consideration
is confirmed by the numerical results of Table \ref{tab:corr_values_b_xj},
which lists the absolute values of the correlation coefficients of
the previously described features with the estimated TOA\ bias for
both the LOS and the NLOS scenarios. From these results it can be
easily inferred that not all the considered features are equally useful
to estimate the TOA bias. For this reason and to simplify our statistical
analysis, we restricted the group of considered features to the set
$\{x_{i,0}$, $x_{i,1}$, $x_{i,2}\}$, which collects the parameters
exhibiting a strong correlation with the bias in the NLOS scenario.

Note that, in principle, the bias $b_{i}$ is not influenced by the
distance $d_{i}$, since it depends only on the thickness of the walls
(or of other obstacles) encountered by the transmitted signal during
its propagation; this is not true, however, for the above mentioned
triple of signal features (see {[}1{]} for further details). Generally
speaking, it is useful to derive a TOA bias estimator which is not
influenced by the transmitter-receiver distance $d_{i}$. In the attempt
of removing the dependence of the features $\{x_{i,0}$, $x_{i,1}$,
$x_{i,2}\}$ on the link distance, we developed the models

\begin{equation}
x_{i,0}=r_{max,i}=r_{max}^{0}-r_{max}^{m}d_{i}\text{,}\label{eq:rmax_signalmodel}
\end{equation}
\begin{equation}
x_{i,1}=\tau_{m,i}=\tau_{m}^{m}d_{i}\label{eq:taum_m_signalmodel}
\end{equation}
and
\begin{equation}
x_{i,2}=\tau_{ds,i}=\tau_{ds}^{0}+\tau_{ds}^{m}d_{i}\label{eq:tauds_m_signalmodel}
\end{equation}
on the basis of our experimental results (and, in particular, on the
basis of the estimates of the joint pdf's $\{f_{b,x_{0}}$, $f_{b,x_{1}}$,
$f_{b,x_{2}}\}$ referring to the three possible couples $(b_{i}$,
$x_{i,j})$ with $j=0$, $1$, $2$); here $r_{max}^{0}$, $\tau_{m}^{m}$
and $\tau_{ds}^{m}$ are random variables%
\footnote{In the rest of the document, the subscript $i$ has been omitted for
simplicity when not strictly necessary.%
}, whereas $\tau_{ds}^{0}$ and $r_{max}^{m}$ are deterministic parameters
having known values. Given these models, the vector of distance-independent
features $\tilde{\mathbf{x}}\triangleq\left[r_{max}^{0},\tau_{m}^{m},\tau_{ds}^{m}\right]^{T}$
can be evaluated from its distance-dependent counterpart $\mathbf{x}=\left[r_{max},\tau_{m},\tau_{ds}\right]^{T}$and
can be used in place of it.

A complete statistical characterization of the estimated bias  and
of the $3$ related signal features we consider is provided by the
joint pdf $f_{b,r_{max},\tau_{m},\tau_{ds}}(\cdot)$ or, equivalently,
by the joint pdf $f_{b,r_{max}^{0},\tau_{m}^{m},\tau_{ds}^{m}}(\cdot)$;
these functions have been estimated from the acquired data, but cannot
be plotted because of the large dimensionality of their domains; for
this reason, we analysed some pdf's deriving from their marginalization.
For instance, Fig. \ref{fig:pdf_2d_bias_taurms} shows the estimated
joint pdf's for the estimated bias $\hat{b}_{i}$ and measured delay
spread for the NLOS and the LOS scenarios; these results deserve the
following comments:
\begin{enumerate}
\item a significant (limited) correlation between these parameters is found
in the NLOS (LOS) case;
\item the null region exhibited by the estimated pdf's is due to the fact
that the TOA bias cannot take on values in the interval $[0,\: t_{wall}/c_{0}]$,
where $t_{wall}$ is the thickness of the wall obstructing the direct
path;
\item large values of the TOA bias are unlikely since they are associated
with small incidence angles of the transmitted signal on the obstructing
wall.

\begin{figure}
\begin{centering}
\includegraphics[width=8cm]{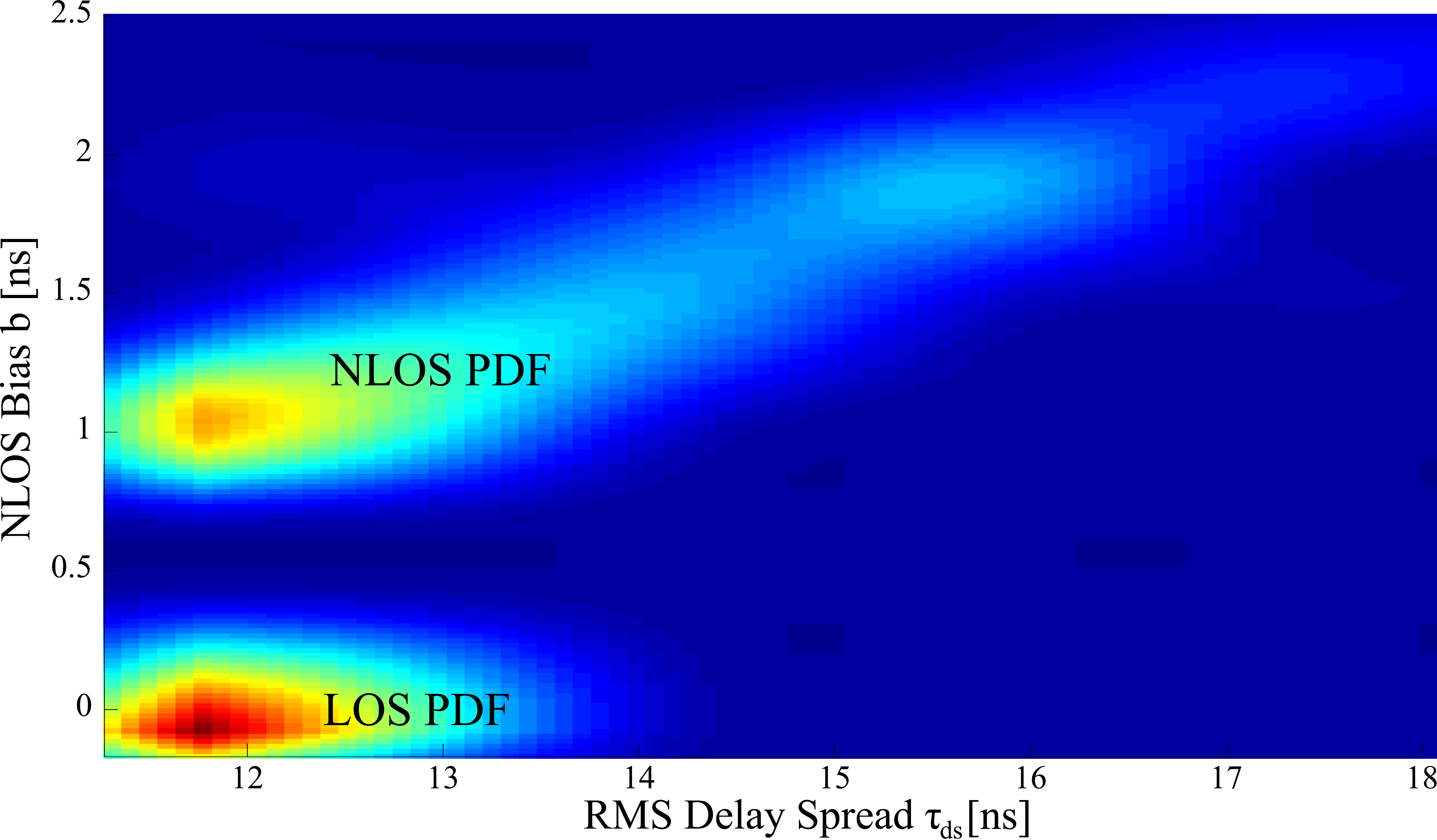}
\par\end{centering}

\caption{Estimated joint pdf of the estimated TOA bias and the delay spread
in NLOS conditions (above) and LOS conditions (below). Note that the
NLOS PDF has been scaled by a factor 3 to improve its visualization.
\label{fig:pdf_2d_bias_taurms}}
\end{figure}

\begin{figure}
\begin{centering}
\includegraphics[width=9cm]{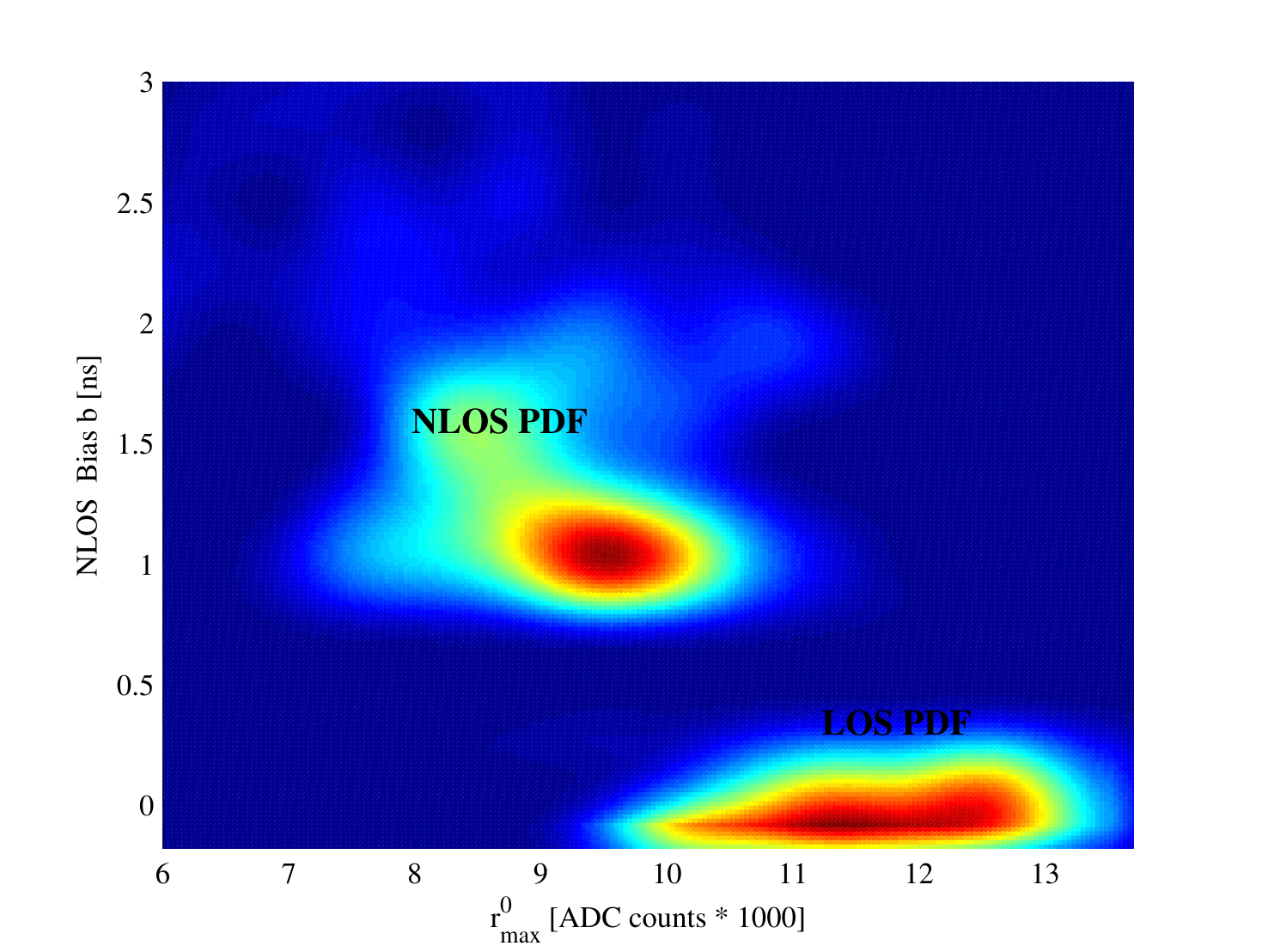}
\par\end{centering}

\caption{Estimated joint pdf of the estimated TOA bias and the $r_{max}^{0}$
random variable in NLOS conditions (above) and LOS conditions (below).
Note that the NLOS PDF has been scaled by a factor 2 to improve its
visualization. \label{fig:pdf_2d_bias_rmax0}}
\end{figure}

\end{enumerate}
Finally, it is important to point out that our experimental data
were also processed to evaluate the correlation coefficient between
the estimated bias $\hat{b}_{i}$ and each of the
distance-independent parameters $r_{max}^{0}$, $\tau_{m}^{m}$ and
$\tau_{ds}^{m}$ (see Table \ref{tab:corr_values_b_rv_indip_distance}
and Fig. \ref{fig:pdf_2d_bias_rmax0}, which represents the joint pdf
of the estimated bias $\hat{b}_{i}$ and the parameter
\textbf{$r_{max}^{0}$}); our results evidence that the last
parameters are less correlated with $b_{i}$ than their distance
dependent counterparts. For this reason, we decided to take into
consideration also the distance-dependent features $\{x_{i,0}$,
$x_{i,1}$, $x_{i,2}\}$ for localization purposes.

\begin{table}
\begin{centering}
\begin{tabular}{|c|c|c|c|}
\hline
\begin{tabular}{c}
Correlation\tabularnewline
with $b_{i}$ \tabularnewline
\end{tabular}  & %
\begin{tabular}{c}
$\tilde{x}{}_{i,0}$\tabularnewline
$r_{max,i}^{0}$\tabularnewline
\end{tabular} & %
\begin{tabular}{c}
$\tilde{x}{}_{i,1}$\tabularnewline
$\tau_{m,i}^{m}$\tabularnewline
\end{tabular} & %
\begin{tabular}{c}
$\tilde{x}{}_{i,2}$\tabularnewline
$\tau_{ds,i}^{m}$\tabularnewline
\end{tabular}\tabularnewline
\hline
\hline
NLOS case & 0.321 & 0.649 & 0.561\tabularnewline
\hline
LOS case & 0.483 & 0.370 & 0.490\tabularnewline
\hline
\end{tabular}
\par\end{centering}

\caption{Absolute values of correlation coefficients between the estimated
bias $\hat{b}_{i}$ and each of the parameters $r_{max}^{0}$, $\tau_{m}^{m}$
and $\tau_{ds}^{m}$. Both LOS and NLOS conditions are considered.\label{tab:corr_values_b_rv_indip_distance}}
\end{table}

\section{Localization algorithms \label{sec:Localization-algorithms}}

\subsection{Introduction}

In this Section we develop various algorithms for two-dimensional
localization in a UWB network composed by $N_{a}$ anchors with \emph{known}
positions $\mathbf{z}_{i}^{a}\triangleq\left[x_{i}^{a},\, y_{i}^{a}\right]^{T}\in\mathbb{R}^{2}$,
$i=0,...,N_{a}-1$ and by a single node (dubbed mobile station, MS,
in the following) with \emph{unknown} position $\boldsymbol{\theta}\triangleq[x,\, y]^{T}\in\mathbb{R}^{2}$.
Any couple of the given $(N_{a}+1)$ devices can operate in a LOS
(NLOS) condition with probability $P_{\text{\tiny{LOS}}}$ ($1-P_{\text{\tiny{LOS}}}$).
The localization algorithms described below try to mitigate the effects
of the NLOS bias error and aim at generating an estimate $\hat{\boldsymbol{\theta}}$
of $\boldsymbol{\theta}$ minimizing the \emph{mean square error}
(MSE) $\mathbb{E}_{\mathbf{\hat{\boldsymbol{\theta}}}}\left\{ ||\mathbf{\hat{\boldsymbol{\theta}}}-\mathbf{\boldsymbol{\theta}||}^{2}\right\} $.
It is also important to point out that localization algorithms developed
for NLOS scenarios usually consist of two steps. In fact, first the
NLOS bias is estimated for each involved link and is used to remove
the bias contribution in the acquired data; then the new data set
is processed by a \emph{least-square} (LS) procedure generating an
estimate of $\mathbf{\boldsymbol{\theta}}$ (e.g., see \cite{nlos_marano},
\cite{decarli}, \cite{venkatesh_buehrer}, \cite{guvenc}). A different
approach, involving implicit estimation of the bias for each link,
is adopted in the following; this approach is motivated by the fact
that the estimation of the bias for the $i$-th link can benefit from
the information acquired from the other ($N_{a}-1$) links; note that
in \cite{nlos_marano} and \cite{guvenc} bias mitigation performed
in a link-by-link fashion is exploited to assign a weight in a \emph{weighted
least-square} (WLS) step, but such an approach is heuristic.

\subsection{Maximum Likelihood Estimation}

If the links between the MS and the $N_{a}$ different anchors are
assumed mutually independent, the ML estimation strategy of the unknown
MS position $\mathbf{\boldsymbol{\theta}}$, given a TOA estimate
and a set of additional signal features for each link, can be formulated
as
\begin{equation}
\mathbf{\hat{\boldsymbol{\theta}}}=\arg\underset{\mathbf{\tilde{\boldsymbol{\theta}}}}{\max}\ln\prod_{i=0}^{N_{a}-1}f_{\mathbf{\tau},\mathbf{x}}(\tau_{i},\mathbf{x}_{i};\mathbf{\tilde{\boldsymbol{\theta}}})\text{.}\label{eq:ml_estimator-1}
\end{equation}
Here, $\mathbf{\tilde{\boldsymbol{\theta}}}=[\tilde{x},\,\tilde{y}]^{T}\in\mathbb{R}^{2}$
denotes the MS trial position, $\tau_{i}$ and $\mathbf{x}_{i}$ are
the TOA and $N_{f}$-dimensional signal vector collecting the received
signal features acquired for the $i$-th link and $f_{\tau,\mathbf{x}}(\cdot,\mathbf{\cdot};\mathbf{\tilde{\boldsymbol{\theta}}})$
is the joint pdf of the TOA and the vector of features parameterized
by the trial position $\mathbf{\tilde{\boldsymbol{\theta}}}$. As
already discussed in Paragraph \ref{sub:Statistical-results}, various
options for $\mathbf{x}$ (and, consequently, for $f_{\tau,\mathbf{x}}(\cdot,\mathbf{\cdot};\mathbf{\tilde{\boldsymbol{\theta}}})$)
are possible; in the following Paragraphs the impact of such options
on the ML strategy (\ref{eq:ml_estimator-1}) is investigated.

\subsubsection{ML estimation strategy for different sets of observed data\label{sub:dimensionality_choice}}

In this Paragraph two different options are considered for the set
of features processed by the ML strategy.

\paragraph{Option A}

In this case it is assumed that the vector of features referring to
the $i$-th link is $\mathbf{x}_{i}=[r_{max,i},\tau_{m,i},\tau_{ds,i}]^{T}$;
this choice is motivated by the large correlation between these random
variables and the link bias $b$ (see Paragraph \ref{sub:Statistical-results}).
Then, the joint PDF $f_{\mathbf{\tau},\mathbf{x}}(\tau_{i},\mathbf{x}_{i};\mathbf{\tilde{\boldsymbol{\theta}}})$
appearing in the ML strategy (\ref{eq:ml_estimator-1}) can be expressed
as (see Eq. (\ref{eq:experimental_sigmodel})):
\begin{equation}
f_{\tau,\mathbf{x}}(\tau_{i},\mathbf{x}_{i};\tilde{\boldsymbol{\theta}})=(f_{b,\tilde{\mathbf{x}}}\otimes f_{w})\left(\tau_{i}-\frac{d_{i}(\mathbf{\tilde{\boldsymbol{\theta}}})}{c_{0}},\tilde{\mathbf{x}}_{i}\right),\label{eq:4d_ml_pdf}
\end{equation}
where (see Eq. (\ref{eq:rmax_signalmodel}),
(\ref{eq:taum_m_signalmodel}), (\ref{eq:tauds_m_signalmodel}))
\begin{equation}
\tilde{\mathbf{x}}_{i}=\left[r_{max,i}+r_{max}^{m}d_{i}(\mathbf{\tilde{\boldsymbol{\theta}}}),\frac{\tau_{m,i}}{d_{i}(\mathbf{\tilde{\boldsymbol{\theta}}})},\frac{\tau_{ds,i}-\tau_{ds}^{0}}{d_{i}(\mathbf{\tilde{\boldsymbol{\theta}}})}\right]^{T}
\end{equation}
$d_{i}(\mathbf{\tilde{\boldsymbol{\theta}}})$ is the distance
between the $i$-th anchor and the MS trial position and
$f_{b,\tilde{\mathbf{x}}}\otimes f_{w}$ denotes the convolution
between the joint pdf $f_{b,\tilde{\mathbf{x}}}(\cdot)$ and the
observation noise pdf
$f_{w}(w)=(2\pi\sigma_{w}^{2})^{-1/2}\exp(-w^{2}/(2\sigma_{w}^{2}))$.

Note that the shape of the function $f_{b,\tilde{\mathbf{x}}}(\cdot)$
under the \emph{hypothesis of} LOS \emph{conditions} ($H_{\text{\tiny{LOS}}}$
event) is substantially different from that found in NLOS conditions
($H_{\text{\tiny{NLOS}}}$ event) \cite{venkatesh_buehrer}; in both
cases this function was estimated applying the procedure described
in \cite{wed_model} to the data collected in our measurement campaign.
This led to two distinct multidimensional histograms, which approximate
the required pdf's $f_{b,\tilde{\mathbf{x}}}\left(b,\tilde{\mathbf{x}}_{i}|H_{\text{\tiny{LOS}}}\right)$
and $f_{b,\tilde{\mathbf{x}}}\left(b,\tilde{\mathbf{x}}_{i}|H_{\text{\tiny{NLOS}}}\right)$
with a certain accuracy depending on: a) the amount of acquired data;
b) the sizes $\Delta b$, $\Delta r_{max}$, $\Delta\tau_{m}$ and
$\Delta\tau_{ds}$ of the quantization bins adopted in the generation
of the histograms. Note that these sizes need to be accurately selected,
since large bins imply a coarse approximation of pdf's, whereas excessively
small bins require a huge amount of data.

Given an estimate of the above mentioned couple of pdf's, the pdf
$f_{b,\tilde{\mathbf{x}}}(\cdot)$ can be evaluated as
\begin{eqnarray}
f_{b,\tilde{\mathbf{x}}}\left(b,\tilde{\mathbf{x}}_{i}\right) & = & P_{\text{\tiny{LOS}}}f_{b,\tilde{\mathbf{x}}}\left(b,\tilde{\mathbf{x}}_{i}|H_{\text{\tiny{LOS}}}\right)+\nonumber \\
 &  & \left(1-P_{\text{\tiny{LOS}}}\right)f_{b,\tilde{\mathbf{x}}}\left(b,\tilde{\mathbf{x}}_{i}|H_{\text{\tiny{NLOS}}}\right)\label{eq:4d_pdf_apriori}
\end{eqnarray}
if the probabilities $P_{\text{\tiny{LOS}}}\triangleq\Pr\{H_{\text{\tiny{LOS}}}\}$
and $P_{\text{\tiny{NLOS}}}\triangleq\Pr\{H_{\text{\tiny{NLOS}}}\}=1-\Pr\{H_{\text{\tiny{LOS}}}\}$
are available, or as
\begin{equation}
f_{b,\tilde{\mathbf{x}}}\left(b,\tilde{\mathbf{x}}_{i}\right)=\frac{1}{2}f_{b,\tilde{\mathbf{x}}}\left(b,\tilde{\mathbf{x}}_{i}|H_{\text{\tiny{LOS}}}\right)+\frac{1}{2}f_{b,\tilde{\mathbf{x}}}\left(b,\tilde{\mathbf{x}}_{i}|H_{\text{\tiny{NLOS}}}\right)\label{eq:4d_pdf}
\end{equation}
if no \emph{a priori} information about the LOS/NLOS conditions are
available.

In principle, evaluating the joint PDF $f_{\mathbf{\tau},\mathbf{x}}(\tau_{i},\mathbf{x}_{i};\mathbf{\tilde{\boldsymbol{\theta}}})$
of the ML strategy entails the computation of the convolution appearing
in (\ref{eq:4d_ml_pdf}); this has not been done in our work, since
it is numerically complicated and a form of implicit smoothing is
already included in the generation of the above mentioned histograms
extracted from the available measurements.

\paragraph{Option B}

In this case the set of features employed in ML estimation consists
of a single element, namely the delay spread (which exhibits the largest
correlation with the NLOS bias; see Table \ref{tab:corr_values_b_xj}),
so that $\mathbf{x}_{i}=\tau_{ds,i}$ and the pdf $f_{\mathbf{\tau},\mathbf{x}}(\tau_{i},\mathbf{x}_{i};\mathbf{\tilde{\boldsymbol{\theta}}})$
of (\ref{eq:ml_estimator-1}) becomes (see (\ref{eq:4d_ml_pdf}))
\begin{equation}
f_{\tau,\mathbf{x}}(\tau_{i},\mathbf{x}_{i};\tilde{\boldsymbol{\theta}})=(f_{b,\tau_{ds}^{m}}\otimes f_{w})\left(\tau_{i}-\frac{d_{i}(\mathbf{\tilde{\boldsymbol{\theta}}})}{c_{0}},\frac{\tau_{ds,i}-\tau_{ds}^{0}}{d_{i}(\mathbf{\tilde{\boldsymbol{\theta}}})}\right)\label{eq:2d_ml_pdf}
\end{equation}
Like in the previous case, the pdf $f_{b,\tau_{ds}^{m}}(\cdot)$ was
estimated in the LOS and NLOS scenarios (see Fig. \ref{fig:pdf_2d_bias_taurms})
from the data acquired in our measurement campaign. Note that this
option leads to a ML localization algorithm which is substantially
simpler than that proposed in the analysis of option A.

\subsubsection{Parameterization of the observations\label{sub:parameterization_choice}}

The ML strategies developed above are based on the joint pdf's $f_{b,r_{max}^{0},\tau_{m}^{m},\tau_{ds}^{m}}(\cdot)$
and $f_{b,\tau_{ds}^{m}}(\cdot)$, which refer to a set of \emph{distance-independent
parameters}. Since the parameters $r_{max}^{0}$, $\tau_{m}^{m}$
and $\tau_{ds}^{m}$ exhibit a lower correlation with $b$ than their
distance-dependent counterparts $r_{max}$, $\tau_{m}$, $\tau_{ds}$
(see Tables \ref{tab:corr_values_b_xj} and \ref{tab:corr_values_b_rv_indip_distance}),
the use of the joint PDF
\begin{equation}
f_{\tau,\mathbf{x}}(\tau_{i},\mathbf{x}_{i};\mathbf{\tilde{\boldsymbol{\theta}}})=(f_{b,\mathbf{x}}\otimes f_{w})\left(\tau_{i}-\frac{d_{i}(\tilde{\boldsymbol{\theta}})}{c_{0}},\mathbf{x}_{i}\right)
\end{equation}
in place of (\ref{eq:4d_ml_pdf}) has been also investigated (similar
comments apply to (\ref{eq:2d_ml_pdf})).

\subsubsection{Estimation of joint pdf's\label{sub:postprocessing_choice}}

As already explained above, the joint pdf's involved in the proposed
ML localization strategies can be easily estimated from the acquired
data using a simple procedure based on dividing the space of observed
data in a set of bins of proper size. Such a procedure generates an
histogram, which, unluckily, entails poor localization performance
if employed as it is, because of the relatively small number of bins
(adopted to avoid empty bins). To mitigate this problem, interpolation
followed by low-pass filtering can be applied to raw experimental
histograms. In particular, we found out that cubic spline interpolation
followed by moving average filtering (with a window size equal to
about 1/15 of the size of the histogram) provides good results (an
example of the resulting pdf is shown in Fig. \ref{fig:pdf_2d_bias_taurms});
the localization strategies adopting this approach are dubbed \emph{interpolated-histogram}
\emph{estimators} in the following.

An alternative to this approach is represented by fitting the
histograms with analytical functions (e.g., polynomials). On the one
hand, this solution offers some advantages with respect to the
previous one, since it requires less memory (only the values of the
coefficients of the fitting equations need to be saved) and provides
better accuracy (the resulting pdf's are smoother). On the other
hand, the use of (multi-dimensional) fitting raises a number of
problems. In fact,
global performance figures, like the \emph{root mean square error}%
\footnote{This parameter is evaluated as the root mean square of the difference
between the values of the original experimental histogram and the
corresponding values generated by a fitting analytical function.%
} (RMSE) are certainly useful, but do not account for the fact that
fitting errors where a pdf takes on small values are less critical
than those in points where the same pdf takes on large values. In
addition, if the dimensionality of the pdf domain is larger than two,
an analytical function generated by a fitting procedure cannot be
easily represented.

In our statistical analysis, the use of polynomial fitting was investigated.
A polynomial degree $N_{pol}=8$ was selected, since this choice ensures
that the resulting RMSE is smaller than $\epsilon_{fitting}=5\cdot10^{-4}$
in both LOS and NLOS scenarios. In the following localization algorithms
employing pdf's generated by polynomial fitting will be dubbed \emph{fitted-histogram}
\emph{estimators}.

\subsection{Iterative Estimation\label{sub:Iterative-Estimator}}

Recently, an iterative estimator of both the channel state (i.e.,
LOS or NLOS conditions) and the NLOS bias $b$ has been proposed in
\cite{venkatesh_buehrer}. Such an estimator relies on the availability
of univariate statistical models for the estimates of TOA, RSS and
RDS; it is shown, however, that estimation accuracy is improved if
the delay-spread is exploited as the only discriminant feature.

\begin{figure}
\begin{centering}
\includegraphics[width=8cm]{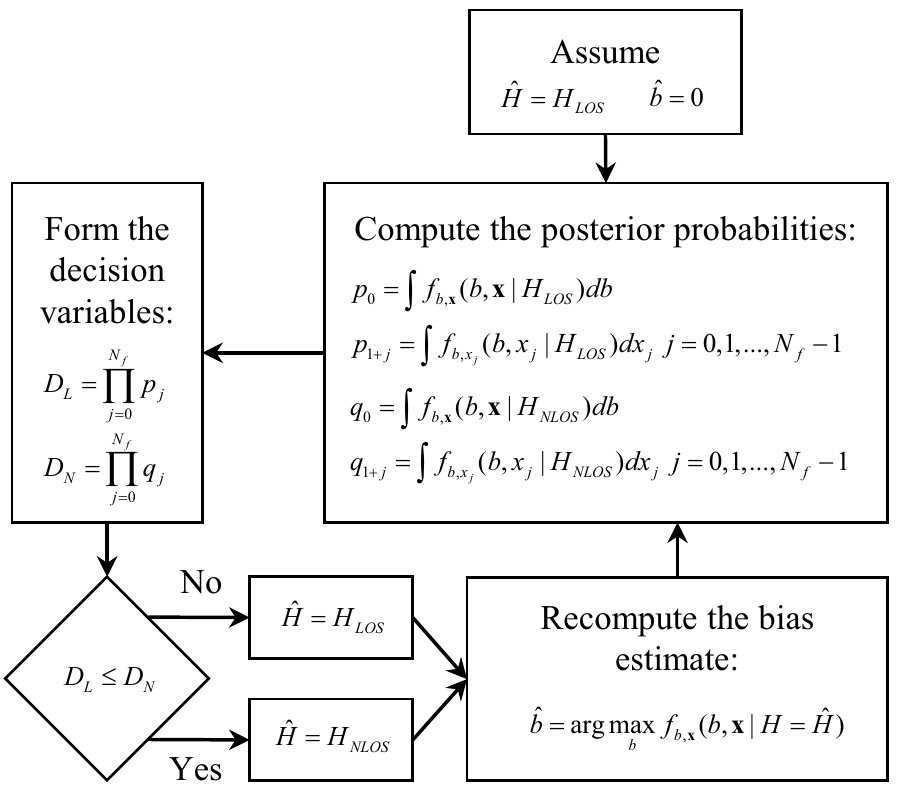}
\par\end{centering}

\caption{Flow diagram of the proposed iterative estimator.\label{fig:iterative_block_diagram}}
\end{figure}

In the following, a modified version of this iterative algorithm
(dubbed \emph{iterative} \emph{estimator}), employing the joint
pdf's described above and extracted from our experimental database,
is proposed. This algorithm is summarized by the flow diagram shown
in Fig. \ref{fig:iterative_block_diagram} and operates as follows.
In its first iteration it starts with a null estimate of the NLOS
bias and, for each link, generates a set of posterior probabilities
integrating the pdf's estimated in LOS and NLOS conditions. Then,
such probabilities are employed to compute a couple of decision
variables on the basis of which a decision $\hat{H}$ on the LOS/NLOS
conditions is taken. Finally, such a decision is exploited to
compute a new bias estimate which is used to start the next
iteration. Note that this algorithm operates in a link-by-link
fashion and that in the diagram of Fig.
\ref{fig:iterative_block_diagram} the dependence on the link index
has been omitted to ease the reading.

\section{Numerical results\label{sec:Numerical-results}}

Our experimental database has also been exploited to assess the RMSE
performance of the proposed algorithms for localization and NLOS bias
mitigation schemes via computer simulations. In our simulations the
MS coordinates are always $(0;0)$; then, following \cite{nlos_marano},
for the $i$-th anchor a received waveform from either the LOS database
(with probability $P_{\text{\tiny{LOS}}}$) or from the NLOS database
(with probability $P_{\text{\tiny{NLOS}}}$) was drawn randomly and
was associated with the position $\mathbf{z}_{i}^{a}=\left(d_{i}\sin(2\pi\frac{i-1}{N_{a}});d_{i}\cos(2\pi\frac{i-1}{N_{a}})\right)$,
where $d_{i}$ is the distance measured for the selected waveform.
Note that these waveforms already include the experimental noise and
thus no simulated noise was imposed on the waveforms (so that the
signal-to-noise ratio is the experimental one). Finally, the parameter
$N_{a}$ has been set to $3$ (worst case which still theoretically
allows unambiguous localization).

\begin{figure}
\begin{centering}
\includegraphics[width=8cm]{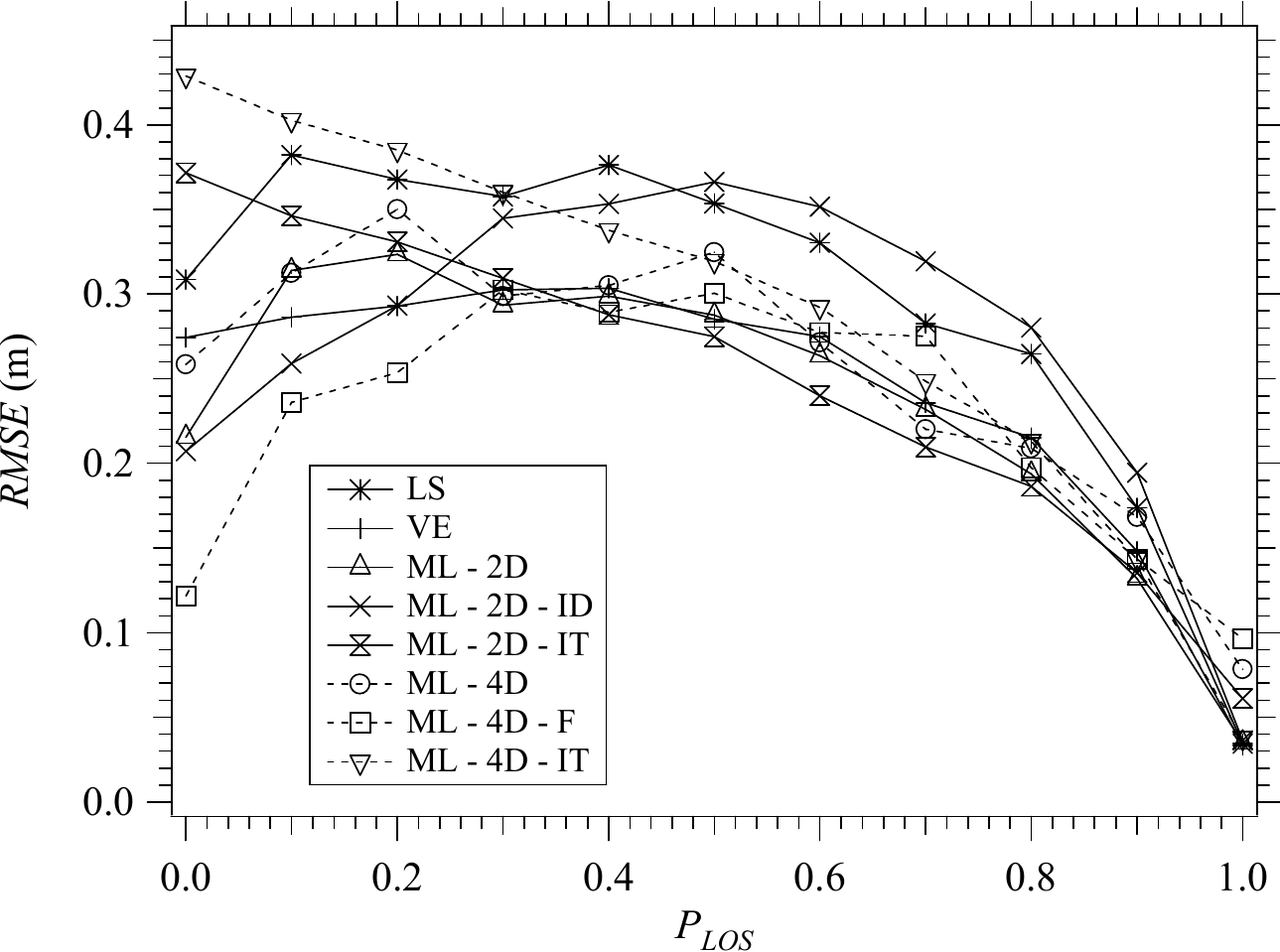}
\par\end{centering}

\caption{RMSE performance versus $P_{\text{\tiny{LOS}}}$ offered by various
localization algorithms.\label{fig:num_results}}
\end{figure}

\begin{figure}
\begin{centering}
\includegraphics[width=8cm]{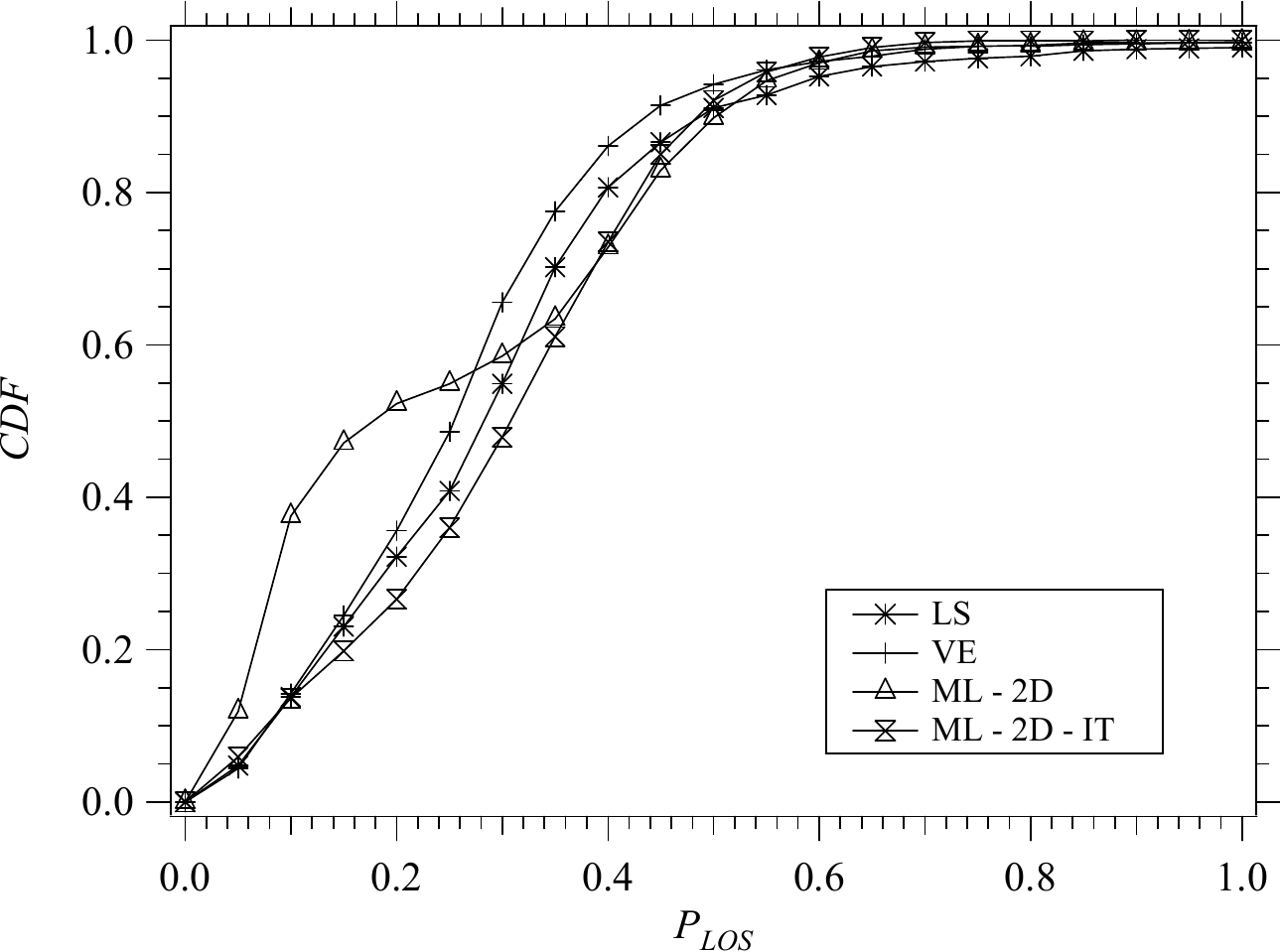}
\par\end{centering}

\caption{CDFs of the RMSE characterizing various localization algorithms when
$P_{\text{\tiny{LOS}}}=0.2$.\label{fig:num_results_cdf}}
\end{figure}

The RMSE performance of the following algorithms has been evaluated:
\begin{enumerate}
\item \textbf{LS} - A standard LS estimator for LOS environments; the estimation
strategy can be expressed as $\hat{\mathbf{\boldsymbol{\theta}}}=\arg\min_{\mathbf{\tilde{\boldsymbol{\theta}}}}\sum_{i=0}^{N_{a}-1}(c_{0}\tau_{i}-d_{i}(\mathbf{\tilde{\boldsymbol{\theta}}}))^{2}$.
\item \textbf{VE} -\emph{ }A LS estimator exploiting TOA measurements corrected
by the algorithm proposed by Venkatesh and Buehrer in \cite{venkatesh_buehrer};
this algorithm relies on a statistical modeling of the propagation
environment based on our experimental database.
\item \textbf{ML-4D} - A ML estimator based on (\ref{eq:4d_pdf}) and employing
an interpolated histogram for the evaluation of its likelihood function
referring to a \emph{distance-dependent} parameterization.
\item \textbf{ML-2D} - A ML estimator based on (\ref{eq:2d_ml_pdf}) and
employing an interpolated histogram for the evaluation of its likelihood
function referring to a \emph{distance-dependent} parameterization.
\item \textbf{ML-2D-ID} - A ML estimator based on (\ref{eq:2d_ml_pdf})
and employing an interpolated histogram for the evaluation of its
likelihood function referring to a \emph{distance-independent} parameterization.
\item \textbf{ML-4D-F} - A ML estimator based on (\ref{eq:4d_pdf}) and
employing a \emph{fitted} histogram for the evaluation of its likelihood
function referring to a \emph{distance-dependent} parameterization.
\item \textbf{ML-4D-IT - }A\textbf{ }LS estimator based on (\ref{eq:4d_pdf})
and exploiting TOA measurements corrected by the modified \emph{iterative}
algorithm illustrated in Section \ref{sub:Iterative-Estimator}.
\item \textbf{ML-2D-IT - }A\textbf{ }LS estimator based on (\ref{eq:2d_ml_pdf})
and exploiting TOA measurements corrected by the modified \emph{iterative}
algorithm illustrated in Section \ref{sub:Iterative-Estimator}.
\end{enumerate}
In estimating the RMSE performance of the ML algorithms listed above
the likelihood functions were always evaluated at the vertices of
a square grid characterized by a step size equal to $10$ mm. Some
numerical results are compared in Fig. \ref{fig:num_results}, which
illustrates the RMSE performance versus the probability $P_{\text{\tiny{LOS}}}$.
These results evidence that:
\begin{enumerate}
\item Different estimators can provide significantly different accuracies
for distinct values of $P_{\text{\tiny{LOS}}}$, when this probability
is not close to unity.
\item The simple LS algorithm is outperformed by all the other algorithms
when $P_{\text{\tiny{LOS}}}\leq0.9$; this is due to the fact that
this strategy does not try to mitigate the NLOS bias.
\item The VE algorithm performs well at the cost of a reasonable complexity,
but offers limited bias mitigation when $P_{\text{\tiny{LOS}}}=0$;
in this case the ML-2D, ML-2D-ID and ML-4D-F estimators perform much
better.
\item The exploitation of a large set of received signal features does not
necessarily allow to achieve better accuracy than a subset of them
(see the curves referring to ML-2D and ML-4D estimators); this is
due to the fact that the correlation between the different couples
of extracted features is typically large, so that they provide strongly
correlated information about the NLOS bias.
\item Distance-dependent parameterization provides better accuracy (see
the curves referring to the ML-2D and ML-2D-ID estimators); this can
be related to the fact that $\tau_{ds}^{m}$ is less correlated with
the NLOS bias than its distance-dependent counterpart $\tau_{ds}$.
\item The ML-4D-F estimator performs better than the ML-4D estimator in
NLOS conditions; this means that the use of fitted histograms entails
an improvement in localization accuracy.
\end{enumerate}
Finally, Fig. \ref{fig:num_results_cdf} shows the cumulative density
function (CDF) of the RMSE localization error characterizing the LS,
VE, ML-2D and ML-2D-IT algorithms when $P_{\text{\tiny{LOS}}}=0.2$.
These results evidence that: a) the RMSE localization error of the
considered algorithms remains below $0.5$ m in average the 80\% of
the cases; b) the ML-2D technique is more accurate than the other
algorithms, in terms of RMSE, for small NLOS errors.


\section{Conclusions\label{sec:Conclusions}}

In this paper various UWB localization techniques processing multiple
features extracted from the received signal to mitigate the problem
of NLOS bias have been described and their accuracy has been assessed
exploiting the experimental data acquired in an measurement campaign.
Our results evidence that: a) a restricted set of features has to
be employed; b) the use of distance-dependent features and of fitted
histograms provide better performance than that offered by distance-independent
features and interpolated histograms.

\section*{Acknowledgements}

The authors would like to thank Prof. Marco Chiani and Prof. Davide
Dardari (both from the University of Bologna, Italy) for lending us
the UWB devices employed in our measurement campaign and the PhD students
Alessandro Barbieri and Fabio Gianaroli for their invaluable help
in our experimental work. Finally, the authors wish to acknowledge
the activity of the Network of Excellence in Wireless COMmunications
(NEWCOM++, contract n. 216715), supported by the European Commission
which motivated this work.

\vspace{5 cm}		
\pagebreak

\begin{IEEEbiographynophoto}{Francesco Montorsi}
received the Laurea degree (summa cum laude) and the Laurea Specialistica degree (summa cum laude) in Electronic Engineering from the University of Modena and Reggio Emilia, Italy, in 2007 and 2009, respectively.

Since 2010 he is a Ph.D. candidate in the ICT PhD school of the University of Modena and Reggio Emilia. Since January 2011 he is a visiting PhD student at the Wireless Communications and Network Science Laboratory of Massachusetts Institute of Technology (MIT). His current research interests include wideband communication systems, indoor localization and inertial navigation systems.

Mr. Montorsi is a member of IEEE Communications Society and served as a reviewer for the IEEE Transactions on Wireless Communications, IEEE Transactions on Signal Processing and several IEEE conferences.
\end{IEEEbiographynophoto}

\begin{IEEEbiographynophoto}{Fabrizio Pancaldi}
was born in Modena, Italy, in July 1978. He received the Dr. Eng. Degree in Electronic Engineering (cum laude) and the Ph. D. degree in 2006, both from the University of Modena and Reggio Emilia, Italy. From March 2006 he is holding the position of Assistant Professor at the same university and he gives the courses of Telecommunication Networks and ICT Systems. He works in the field of digital communications, both radio and powerline. His particular interests lie in the wide area of digital communications, with emphasis on channel equalization, statistical channel modelling, space-time coding, radio localization, channel estimation and clock synchronization.
\end{IEEEbiographynophoto}

\begin{IEEEbiographynophoto}{Giorgio Matteo Vitetta}
(S'89, M'91, SM’99) was born in Reggio Calabria, Italy, in April 1966. He received the Dr. Ing. Degree in Electronic Engineering (cum Laude) in 1990 and the Ph. D. degree in 1994, both from the University of Pisa, Italy. In 1992/1993 he spent a period at the University of Canterbury, Christchurch, New Zealand, doing research for digital communications on fading channels. From 1995 to 1998 he was a Research Fellow at the Department of Information Engineering of the University of Pisa. From 1998 to 2001 he has been holding the position of Associate Professor of Telecommunications at the University of Modena and Reggio Emilia. He is now Full Professor of Telecommunications in the same university.  His main research interests lie in the broad area of communication theory, with particular emphasis on coded modulation, synchronization, statistical modeling of wireless channels and channel equalization.  He is serving as an Editor of both the IEEE Transactions on Communications (Editor for Channel Models and Equalization in the Area of Transmission Systems) and the IEEE Transactions on Wireless Communications (in the Area of Transmission Systems).
\end{IEEEbiographynophoto}

\vspace{5 cm}		
\pagebreak

\end{document}